\documentclass[prl,epsf,twocolumn]{revtex4}
\usepackage{amsfonts}
\usepackage{graphicx}
\usepackage{graphics}
\usepackage{eepic,epsfig}
\usepackage{dcolumn}
\usepackage{bm}

\newcommand{\nn}{\nonumber}
\begin{document}
\title{Localization-delocalization transition in a presence
of correlated disorder: The random dimer model}
\author{Tigran Sedrakyan}
\email{tigrans@moon.yerphi.am} \affiliation{ITP, University of
Bern, Sidlerstrasse 5, CH-3012, Bern, Switzerland and Yerevan
Physics Institute, Br. Alikhanian str.2, Yerevan 36, Armenia}
\date{\today}
\begin{abstract}
The one dimensional dimer model is investigated and the
localization length calculated exactly. The presence of
delocalized states at $E_c = \epsilon_{a,b}$ of two possible
values of the chemical potential in case of
$\mid\epsilon_a-\epsilon_b\mid \leq 2 $ is confirmed and the
corresponding indices of the localization length were calculated.
The singular integral equation connecting the density of states
with the inverse of the localization length  is solved and the
analytic expression for the density of states compared with the
numerical calculations.
\end{abstract}
\pacs{......}
\maketitle

\newpage
\pagestyle{plain}

 After Anderson's works \cite{And58}, \cite{AbouAndThou} it
became clear that all states of the systems in one or two
dimensional spaces, which are putted into the fully disordered
potential field (independent site-energy disordered Anderson
model), are exponentially localized. In \cite{Eva4}-\cite{Eva5} it
has been shown, that this claim is true for any strength of
disorder and the localization is present in less than three
dimensions even for an infinitesimal amount of full disorder. The
idea arises that the type of disorder underlies of
localization-delocalization (insulator-metal) transition of one
dimensional systems and in order to understand its nature we need
to investigate the conditions, under which delocalized states can
appear.
  That is why it is necessary to consider
experimentally \cite{SS21} and theoretically \cite{SS3}- \cite{SS}
one-dimensional systems with short(or long)-range correlated
disorder, where the random variables are not fully independent,
but are correlated  at short (long)  distances. The vanishing of
the localization and the appearance of diffusion of electrons by
correlations was further put forward for the explanation of high
conductivity of polymers such as doped polyaniline, which can be
approached by random-dimer model \cite{7SS}, \cite{SS6}. The
transport properties of random semiconductor superlattices also
exhibit \cite{P-D, SS, SS21} delocalization in case of correlated
disorder.

One of the simplest tight-binding and numerically best studied
model with the nearest-neighbor correlated disorder is the
random-dimer model \cite{DWP}-\cite{SS9}, where one (or both) of
the two possible values of site potentials $\epsilon_a$ and
$\epsilon_b$ are random in pairs and appearing with probabilities
$p$ and $1-p$. In these papers authors have analyzed the mentioned
model and it was shown numerically that initially localized
electron can become delocalized if $|\epsilon_a-\epsilon_b|\leq
2t$ (here $t$ is a constant electron hopping coefficient). In
 the paper \cite{Eva} the authors have studied the dimer model by numerical
and semi-analytical methods. In case when both site potentials
appear in pairs, they have calculated the critical energies
coinciding with results of \cite{DWP}, and correlation length
index $\nu=2$ (superdiffusion) when $|\epsilon_a-\epsilon_b|<2t$,
and equal to 1 for $|\epsilon_a-\epsilon_b|=2t$. They present also
some calculations of the density of states and the correlation
length indices for different values $p$ for the probability. In
the papers \cite{DGK1}, \cite{DGK2} similar results were obtained
by use of numerical methods. An interesting analytic approach was
developed in \cite{IK}.

Since in all papers above the calculations were done mainly
numerically (though there are some analytical arguments supporting
the presence of delocalized point), the question whether the
correlation length is infinite at some point (critical point) or
finite and equal to lattice size (extended state), remains open.
In order to answer this questions it is reasonable to calculate
the critical energy, the correlation length, the correlation
length index and the density of states of paired-dimer model {\it
analytically} and compare them with already obtained numerical
results. Our aim in this article is to fill this gap and, by use
of technique developed in \cite{P-D} and \cite{SS}, investigate
the physical quantities mentioned above analytically.

In this article we will concentrate on one-dimensional
tight-binding model of random binary alloy with on site potential
field $\epsilon_a$, $\epsilon_b$, which are assigned randomly on
the lattice sites with probabilities $p$ and $(1-p)$
correspondingly. As it was mentioned above, the diffusion of
electrons may occur if we introduce a short range correlation in
the distribution of the site potentials. Here we focus on a
particular realization of the dimer model, when the site
potentials appear always in pairs. The example of certain lattice
segment is
$...\epsilon_a\epsilon_a\epsilon_a\epsilon_a\epsilon_b\epsilon_b\epsilon_a
\epsilon_a\epsilon_b\epsilon_b\epsilon_b\epsilon_b
\epsilon_a\epsilon_a...$, and it is clear that we have correlation
in the probability distribution of nearest-neighbor site
potentials.

We calculate the dimensionless Landauer resistance(correspondingly
Landauer exponent) analytically and compare it with numerical
simulations of Lyapunov exponent. In articles \cite{ATAF} Anderson
and coauthors have argued, that the Lyapunov exponent is simply
the half of Landauer one. Later, Kappus and Wegner have shown in
\cite{KW} (confirmed in \cite{DG}), that the band center can be
anomalous. In Figure 1 we present our result (for
$\epsilon_a-\epsilon_b= t$ case) for the Lyapunov and the half of
Landauer exponents. One can see the coincidence in a wide central
region (and in the whole region in general) demonstrating that the
Lyapunov and Landauer exponents have a same critical behavior.
Exact analytic calculations of the Landauer resistance and the
correlation length shows the existence of two real critical points
with critical energies $E_{crit}^{(1)}=\epsilon_a$, and
$E_{crit}^{(2)}=\epsilon_b$ if $|\epsilon_a-\epsilon_b|\leq 2t$.
All other states are localized. We confirm the results of
\cite{DWP} and \cite{Eva} in the region
$|\epsilon_a-\epsilon_b|<2t$ where there is a super-diffusion with
the correlation length index $\nu=2$. However the situation is
different in case $\epsilon_a-\epsilon_b=2t$. Our analytical
calculations show, that the correlation length index $\nu=1$ when
approaching to the critical points $\epsilon_a$ and $\epsilon_b$
from inside, while  $\nu=1/2$, when approaching from outside of
the segment [$\epsilon_a$,$\epsilon_b$]. It seems to us that this
fact was not observed in the earlier works.

We have analyzed also the density of states. In article \cite{T}
D.Thouless has found a singular integral equation connecting the
density of states with the Lyapunov exponent. By differentiating
this equation(which is Carleman´s equation) one can reduce it to
Hilbert transform problem. In order to find a solution by
expressing the density of states via the derivative of the
Lyapunov exponent we have used the theory of singular integral
equations presented in the book by Muskhelishvili \cite{M}. In a
class of functions which have a finite derivatives in all points
the solution is defined by one arbitrary constant, which will be
fixed by the condition that the integral of density of states is
equal to one. We are presenting this solution with the use of half
of Landauer exponent and compare it with the numerical calculation
(which have been also presented earlier in \cite{Eva}). Again, we
found an excellent   correspondence of the two results in the
central critical region. The numerical data exhibits strong
 fluctuations at the edges of the energy region due to luck of enough large
 $N$ size of thermodynamic limit. But if one will
allow the presence of points where the derivative of the density
of states is infinite, then one can find other solutions.
Unfortunately present numerical calculations are not allowing to
answer to this question precisely, but it is clear, that by fixing
the number and places of the singular points we can find a unique
solution for the density of states.

The Schr\"{o}dinger equation for the stationary eigenstates
$\psi_i(E)$ of the eigenenergy $E$ is
\begin{equation}
(E-\epsilon_i)\psi_i -\psi_{i+1}-\psi_{i-1}=0, \quad i=1,2,\ldots, N,
\label{H}
\end{equation}
where  $\epsilon_n$ is the chemical potential at the site $n$ (it
can be regarded also as an external potential), $N$ is the number
of atoms in the system and we have re-scaled the energies by the
hopping parameter $t$. Let us define
$\epsilon_{a}-\epsilon_{b}=m$.

The Schr\"{o}dinger equation~(\ref{H}) can be written via the
$2\times 2$ Transfer matrix $T_i$ as follows
\begin{equation}
\left(
\begin{array}{l}
\psi_{i+1} \\
\psi_{i}
\end{array}
\right) = \left(
\begin{array}{cc}
E - \epsilon_i & -1 \\
1 & 0
\end{array}
\right) \left(
\begin{array}{l}
\psi_{i} \\
\psi_{i-1}
\end{array}
\right) \equiv T_{i} \left(
\begin{array}{l}
\psi_{i} \\
\psi_{i+1}
\end{array}
\right).  \label{promotion}
\end{equation}
One can easily find out following property $T_i^{-1}=
\sigma_2 T_i^{\dag}\sigma_2$
($\sigma_2$ is Pauli matrix),
which means that $T_i$ is an element
of the group $SU(1,1)$.
By iterating this equation we can relate $(\psi_{i+1},\psi_{i})$ and $%
(\psi_{0},\psi_{1})$:
\begin{equation}
\left(
\begin{array}{l}
\psi_{i+1} \\
\psi_{i}
\end{array}
\right) = \prod_{k=i}^{1} T_{k} \left(
\begin{array}{l}
\psi_{1} \\
\psi_{0}
\end{array}
\right) \equiv M_{i} \left(
\begin{array}{l}
\psi_{1} \\
\psi_{0}
\end{array}
\right),  \label{transfer}
\end{equation}

where the product $M_{N}=\prod_{i=N}^{1}\,T_i$ is the total
Transfer matrix of the system of $N$ unit cells.

Oseledec's theorem~\cite{O} states that the eigenvalues of following
matrices
\begin{equation}
\Gamma_N= \left(M_{N} M_{N}^{\dag}
\right)^{1/{2N}}, \label{G}
\end{equation}
have a limit $e^{\bar\gamma}$ with nonnegative
$\bar\gamma$, which called Lyapunov exponent. Due to
self-averaging property $\bar\gamma$ is equal to $<\log \Gamma >$.
This quantity is very hard to calculate directly.
Instead, the calculation of the matrix $G_N =M_{N}\otimes
M_{N}^{\dag}$ and its disorder average can be done exactly
\cite{P-D}. The dimensionless Landauer resistance
$\rho=<G_{12,21}>$ is defined by  ${12, 21}$ matrix element of
$G$.

By use of the formula for decomposition of direct product of two
spin-1/2 states into the direct sum of scalar and spin-1 states,
we can decompose the direct product of $T_j$ and $T_j^{-1}$ matrices
into scalar and spin-1 parts as
\begin{equation}
(T_j)^{\alpha}_{\alpha^{\prime}}(T_{j}^{-1})^{\beta^{\prime}}_{\beta}= {%
\frac{1 }{2}}(\delta)^{\alpha}_{\beta}
(\delta)^{\beta^{\prime}}_{\alpha^{\prime}}+ {\frac{1 }{2}}%
(\sigma^{\mu})^{\beta^{\prime}}_{\alpha^{\prime}}
\Lambda_{j}^{\mu\nu}(\sigma^{\nu})^{\alpha}_{\beta},  \label{TT1}
\end{equation}
where $\sigma^{\nu}, \;\; \nu=1,2,3$ are Pauli matrices. In this expression the
Kronecker $\delta$'s define the scalar part, while
\begin{equation}
\Lambda_{j}^{\mu\nu}={\frac{1}{2}}\,{\rm Tr}\left(T_j\sigma^{\mu}T_{j}^{-1}
\sigma^{\nu}\right)  \label{L}
\end{equation}
defines the spin-1 part.
By multiplying the expression~(\ref{TT1}) from
the left and right hand sides by $\sigma_2$ we will have
\begin{equation}  \label{TT2}
(T_j)^{\alpha}_{\alpha^{\prime}}(T_{j}^{+})^{\beta^{\prime}}_{\beta}=
{\frac{1}{2}}(\sigma_2)^{\alpha}_{\beta}
(\sigma_2)^{\beta^{\prime}}_{\alpha^{\prime}}+ {\frac{1 }{2}}(\sigma^{\mu}
\sigma_2)^{\beta^{\prime}}_{\alpha^{\prime}}
\Lambda_{j}^{\mu\nu}(\sigma_2 \sigma^{\nu})^{\alpha}_{\beta}.
\end{equation}

As it was shown in article \cite{P-D}, the similar expression
is correct for the total Transfer matrix $M_N$, but instead of
$\Lambda_{j}^{\mu\nu}$ we will have a product $\prod_{j=1}^N
\Lambda_{j}$.

Now we should take into account the disorder and calculate the average of
$\Gamma$ by random distribution of pairs of potentials
$(\epsilon_a,\epsilon_a)$
and $(\epsilon_b,\epsilon_b)$.
\begin{equation}
\langle G \rangle ={\frac{1 }{2}} \sigma_2 \otimes \sigma_2 +
{\frac{1}{2}} (\sigma^{\mu} \sigma_2)\otimes (\sigma_2
\sigma^{\nu}) \left(\prod_{j=1}^{N/2}
\langle\Lambda_{j}^2\rangle\right)^{\mu \nu}, \label{Gamma}
\end{equation}
where $3\times 3$ matrix $\Lambda_j^2$ is defined as
\begin{eqnarray}
\label{L2}
(\Lambda_{j}^2)^{\mu\nu}={\frac{1}{2}}\,{\rm Tr}\left(T_j^2\sigma^{\mu}
[T_{j}^2]^{-1}
\sigma^{\nu}\right).
\end{eqnarray}

In the Dimer model under consideration we should average the
square of $\Lambda_i$ as
\begin{eqnarray}
\label{L22} \Lambda =\langle\Lambda_j^2\rangle =
p\Lambda_j^2(\epsilon_a)+(1-p)\Lambda_j^2(\epsilon_b),
\end{eqnarray}
where the $\Lambda_j(\epsilon_a)$ and $\Lambda_j(\epsilon_b)$ are
calculated for the site potentials $\epsilon_{a}$ and $\epsilon_{b}$
respectively.

Without loosing the generality one can choose $\epsilon_a=-m/2$
and  $\epsilon_b=m/2$. Then the resulting matrix $\Lambda$ will have the
following elements:
\begin{eqnarray}
\label{LL}
\Lambda^{11} &=&
[2+((E-m/2)^2-4)(E-m/2)^2]\frac{1-p}{2}\nn\\
&+&
[(E+m/2)^4-4(E+m/2)^2+2]\frac{p}{2}\;,  \nonumber\\
\Lambda^{12}  &=&
((E-m/2)^2-2)(E-m/2)^2\frac{1-p}{2}\frak{i}\nn\\
&+&
(E+m/2)^2((E+m/2)^2-2)\frac{p}{2}\frak{i}\;,
\nonumber\\
\Lambda^{13}  &=&
((E-m/2)^2-2)(E-m/2)(p-1)\nn\\
&-&(E+m/2)((E+m/2)^2-2)p \;,  \nonumber\\
\Lambda^{21}  &=&
((E-m/2)^2-2)(E-m/2)^2\frac{p-1}{2}\frak{i}\nn\\
&-&
(E+m/2)^2((E+m/2)^2-2)\frac{p}{2}\frak{i}\;,
\nonumber\\
\Lambda^{22}  &=&
((E-m/2)^4+2)\frac{1-p}{2}+((E+m/2)^4+2)\frac{p}{2}\;, \nonumber\\
\Lambda^{23}  &=&
(E-m/2)^3(1-p) \frak{i}+(E+m/2)^3p \frak{i}\;, \nonumber\\
\Lambda^{31} &=&
((E-m/2)^2-2)(E-m/2)(1-p)\nonumber\\
&+&(E+m/2)((E+m/2)^2-2)p\;, \nonumber\\
\Lambda^{32} &=&
(E-m/2)^3(1-p) \frak{i}+(E+m/2)^3p \frak{i}\; ,\nonumber\\
\Lambda^{33} &=&
1-2(E+m/2)^2+4(E+m/2)m(1-p)\nonumber\\
&-&2m^2(1-p) \;.
\end{eqnarray}
By the formula (\ref{transfer}) $\psi_N=M_{12}\psi_0$ we have
$\frac{\mid \psi_N^2\mid}{\mid \psi_0^2\mid} = M_{12}M_{21}^{+}$
and from
\begin{eqnarray}
\label{MMM}
\langle M\otimes M^+\rangle_{12;21}&=& \frac{1}{2}[({\Lambda}^{
N/2})^{22}-({\Lambda}^{N/2})^{11} -i ({\Lambda}^{N/2})^{12}\nonumber\\
&+&i ({\Lambda}^{N/2})^{21}] \simeq\lambda_{max}^{N/2}
\end{eqnarray}
it is clear, that
$\langle \frac{\mid\psi_N\mid ^2}{\mid\psi_0\mid ^2}\rangle \sim
\lambda_{max}^{\frac{N}{2}}
 = e^{2 N \gamma}=e^{\frac{2 N}{\xi(E)}}$,
where
$\lambda_{max}=e^{4 \gamma}=e^{\frac{4}{\xi(E)}}$
is the closest to unity eigenvalue of $\Lambda=\langle
\Lambda^2_j\rangle$. Therefore the quantity $\rho=\langle M\otimes
M^+\rangle_{12;21}$, which is (see for example \cite{T}) nothing
but inverse module square of the Green function, defines the ratio
of retarded over transmitted probabilities $\frac{\mid r
\mid^2}{\mid t \mid^2}$ and can be regarded as the Landauer
resistance. The closest to unity from above eigenvalue of $\langle
\Lambda_j^2\rangle$ for an energy value $E$ defines  the
localization length $\xi=\frac{4}{ln\lambda_{max}}$  given in
units of the length of the unit cell.

Strictly speaking the Lyapunov exponent $\bar\gamma$, defined by
(\ref{G}), differs from the Landauer $\gamma$. According to
Anderson et al. \cite{ATAF} $\bar\gamma= \gamma/2$. But, even when
according to \cite{KW, DG} there is an anomaly, it is absolutely
clear, that both exponents should define the same critical
behavior(with the same critical index) at the same point $E_c$.
Around that point they can differ only by the constant
multiplicative factor, which is probably close to 1/2. In Figure
1. we demonstrate the result of numerical simulations of Lyapunov
exponent (dots), calculated iteratively in a standard way
\cite{MK}, and the half of Landauer exponent, which one can get
from the formulas (\ref{LL}) and (\ref{MMM}). We see perfect
coincidence in the wide region around band center and in whole
region in general.
\begin{figure}
\includegraphics{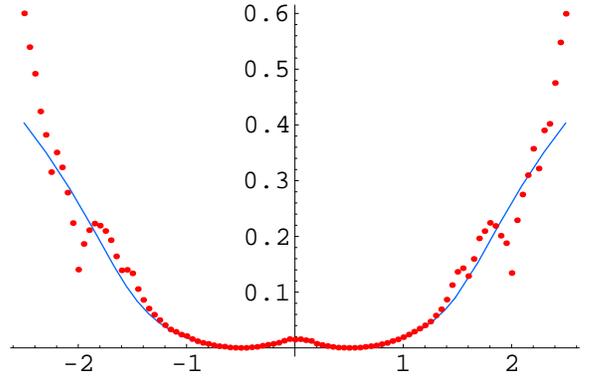}
\caption{m=1, p=1/2. Dots mark the Lyapunov exponent obtained by
simulation of the chain of length 30000, while the curve is the half of
Landauer exponent.}
\label{Fig1}
\end{figure}

The delocalized states are corresponding to the critical points
$\xi(E_c)=\infty $ and can be found by the condition
\begin{eqnarray}
&&det\mid
id-\Lambda \mid =\\
&-&2(E+m/2)^2(E-m/2)^2m^2(1-p)p
=0\nn
\end{eqnarray}
and we see that $E_c=\pm m/2$, which is in full accordance with
the original numerical observations of the articles
\cite{SS3,SS5,SS6,7SS}.

One can find out the closest to unity eigenvalue $\lambda_{max}$
by solving characteristic cubic equation for the matrix (\ref{LL})
and, therefore, calculate exactly the correlation length
$\xi(E)=4/\log[\lambda(E)]$.

For future considerations we will set $p=1/2$ for simplicity.
 Asymptotic-s of $\xi(E)$ around critical
points $E_c=\pm m/2$,  which defines the correlation length
indices $\nu=2$, are $\xi(E) = 6/(E+m/2)^2$, (or $6/(E-m/2)^2$),
for $m < 2$
and
\begin{eqnarray}
\label{A}
 \xi(E) =\left\{\begin{array}{l}
\sqrt2(E+1)^{-1/2}\qquad \;{\text for} \; E\rightarrow -1^-\\
4(E+1)^{-1}\qquad \qquad {\text for} \; E\rightarrow -1^+\\
4(E-1)^{-1}\qquad \qquad {\text for} \; E\rightarrow 1^-\\
\sqrt2(E-1)^{-1/2}\qquad \;{\text for} \; E\rightarrow 1^+
\end{array}
\right.
\end{eqnarray}
for $m=2$ respectively.

As we see, $\nu=2$ in the case $m<2$. For $m=2$ we see $\nu= 1/2$,
when we are approaching the critical points $E_c=\pm 1$ from the
outside of the region $[-1,1]$, while $\nu=1$, when we are
approaching them from the inside. One can check by direct
calculations that these critical indices are independent from the
choice of value of $p$.

Now let us analyze the density of states. According to article
\cite{T} the density of states $\rho(E)$ are connected with the
Lyapunov exponent $\bar\gamma(E)$ as
\begin{eqnarray}
\label{Carleman} \bar\gamma(E)=
\int_{\cal{D}}\hspace{-0.55cm}-\rho(E^\prime)\log[E^\prime -E]d
E^\prime ,
\end{eqnarray}
where $\cal{D}$ is the energy region of non-zero density of states
and $\int_{\cal{D}}\hspace{-0.55cm}- \;$ should be understood as a
Cauchy principal value. In our problem of dimers it is defined by
the interval $-2-m/2 < E < 2+m/2$, because the kinetic energy on
lattice can vary only in the interval $[-2, 2]$.

Let us now differentiate the left and right hand sides of the
equation (\ref{Carleman}) by $E$ and bring it to Hilbert`s
transform form
\begin{eqnarray}
\label{Hilbert} \frac{d\bar\gamma(E)}{d E}=
 - \int_{\cal{D}}\hspace{-0.5cm}- \rho(E^\prime)\frac{1}{E^\prime -E}
 d E^\prime.
\end{eqnarray}

The formula (\ref{A}) shows that the cases $m=2$ and $m < 2$ are
essentially different. Therefore we will consider cases  $m=1$ and
$m=2$ separately.

 In case of $m=1$, when we suppose that all
derivatives of the $\rho(E)$ and $\gamma^{\prime}(E)$ functions
are finite in the energy interval [-2.5, 2.5] beside the end
points, according to \cite{M} the general solution will be
\begin{eqnarray}
\label{solution1} \rho(E)&=&
\frac{1}{\pi\sqrt{2.5^2-E^2}}\\
&\cdot&\left(\frac{1}{\pi}
\int_{-2.5}^{2.5}\hspace{-0.9cm}-\sqrt{2.5^2-(E^\prime)^2}
\frac{d\bar\gamma(E^\prime)/dE^\prime} {E-E^\prime} d E^\prime + C
\right).\nn
\end{eqnarray}
It appears that $\int_{\cal{D}} \rho(E) d E = C$ and $C$ therefore
should be set to be equal to one.

Since the Lyapunov exponent is hard to calculate analytically
exactly lets us, following articles \cite{ATAF}, put the half of
Landauer exponent $\gamma(E)/2=\frac{ln\lambda_{max}}{8}$ into the
expression (\ref{solution1}) and find the corresponding density of
states. In the Fig.2 we present this solution for the case of
$m=1$ together with the numerical calculation of the density of
states, made for a chain of length $N=100$, averaged over random
dimer $\epsilon$-s of the ensemble of 20000 samples. The numerical
calculations were made by exact diagonalization of the
Hamiltonian.  As in the case of exponents (see Fig.1) we see very
good correspondence of two curves in the energy region $E \in
(-1.8, 1.8)$, but numerical dates strongly fluctuate outside of
that. This is the region, where the correlation length becomes of
order of lattice size and where the Lyapunov exponent differs from
the Landauer one. As it follows from the theory of singular
integral equations \cite{M} in case when $\rho(E)$ is infinite at
some points one should divide whole energy band into segments with
singular(and zero) points at the ends and write corresponding
solution of the the Hilbert`s problem \ref{Hilbert}. For
simplicity we represent in Fig.2 the solution for case of absence
singular points. But if one will be able to consider longer chain
and observe in the numerical curve of $\rho(E)$ presence of
singular points (one can see indication of that in Fig.2) then,
following \cite{M}, it will be possible to organize better fit of
the numerical dates. It will not affect the good agreement of
numerical simulations and analytic calculations on the basis of
Landauer exponent in the most important central critical region.
\begin{figure}
\includegraphics{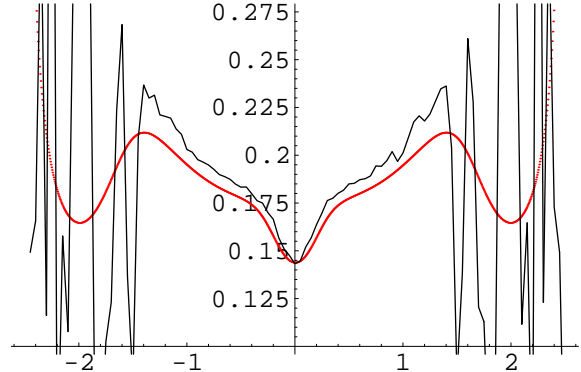}
\caption{The density of states versus energy for the case m=1.
Smooth curve represents the solution of the singular integral
equation, while the fluctuating curve represents the numerical
simulation for the ensemble of 20000 chain`s of length
$N=100$.}\label{Fig2}
\end{figure}

 The case of $m=2$ is more complicated. One can easily find out
from  expression (\ref{A}), that the Landauer exponent
$\gamma^{\prime}(E)$ has singular derivative at the points $E=\pm
1$ approaching to them from the outside. Therefore, first we
should extract singular part from the $\gamma^{\prime}(E)$.
Analyze shows, that the function
$\chi(E)=\gamma^{\prime}(E)-\eta(E)$ with
\begin{eqnarray}
\label{eta} \eta(E)=\left\{
\begin{array}{l}
-\frac{2}{\sqrt{E^2-1}}, \qquad \qquad for E\leq -1\nn\\
0,\qquad \qquad \qquad  \quad for -1\leq E \leq 1\nn \\
\frac{2}{\sqrt{E^2-1}}, \qquad \qquad \; for 1\leq
\end{array}\right.
\end{eqnarray}
has no singular point. Therefore we can find the Hilbert transform
for $\chi(E)$, add to it the Hilbert transform of $\eta(E)$ and
obtain $\rho(E)$. Moreover, numerical calculations show also
presence of some other singular points, in the region of energies
$[-3,-1]\bigcup[1,3]$. Fig.3 shows that the points $\pm2, \pm2.4$
looks singular, therefore, following the technique of solving
singular integral equations presented in book \cite{M} for the
case of multiply connected region of integration, one can find the
following expression for the density of states
\begin{widetext}
\begin{eqnarray}
\label{solution2} &&\rho(E)=
\frac{\sqrt{E^2-1.3^2}\sqrt{E^2-2.05^2}\sqrt{E^2-2.45^2}}{\pi\sqrt{E^2-3^2}
\sqrt{E^2-2.4^2}\sqrt{E^2-2^2}\sqrt{E^2-1}}\\
&&\cdot\left(\frac{1}{\pi} \int_{\cal{D}}\hspace{-0.5cm}-
\;\;\;\frac{\sqrt{(E^\prime)^2-3^2}\sqrt{(E^\prime)^2-2.4^2}
\sqrt{(E^\prime)^2-2^2}\sqrt{(E^\prime)^2-1}}
{\sqrt{(E^\prime)^2-1.3^2}\sqrt{(E^\prime)^2-2.05^2}
\sqrt{(E^\prime)^2-2.45^2}}\frac{d\bar\gamma(E^\prime)/dE^\prime}
{E-E^\prime} d E^\prime + \frac{i}{2} \right)
 +\frac{1}{2 \pi \sqrt{1-E^2}}
\;\theta[E;-1,1],\nn
\end{eqnarray}
\end{widetext}
where $\cal D$ $= [-3,-2.45]\bigcup[-2.4,-2.05]\bigcup[-2,-1.3]
[-1,1]\\ \bigcup[1.3,2]\bigcup[2.05,2.4]\bigcup[2.45,3]$ and the
function $\theta[E;-1,1]$ defined as follows
\begin{eqnarray}
\label{theta} \theta[E;-1,1]=\left\{\begin{array}{l} 1, \qquad if
\qquad E \in
[-1,1]\nonumber\\
0, \qquad if \qquad E \not\in [-1,1].
\end{array}
\right.
\end{eqnarray}
This term appears in a result of Hilbert transform of $\eta(E)$.

\begin{figure}
\includegraphics{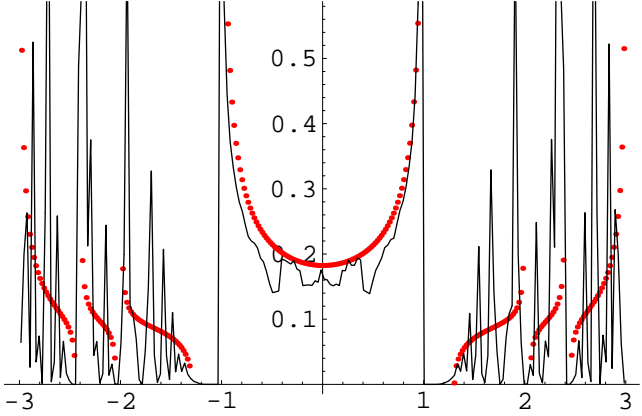}
\caption{The density of states versus energy for the case m=2.
Dots represent the solution of the singular integral equation,
while the fluctuating curve represents the numerical simulation
for the ensemble of 20000 chain`s of length $N=100$.}\label{Fig3}
\end{figure}
In the Fig.3 we present this solution defined by the half of
Landauer exponent together with numerical results for $N=100$ and
20000 iterations. Since the half of Landauer exponent is in good
agreement with the Lyapunov exponent in the center of energy band
we expect good agreement between numerical and analytic results
for the density of states there. Indeed, as in case of $m=1$ we
see, that the analytic curve fits the numerical results in the
region $E \in [-1,1]$ with 1\% accuracy. We see also strong
fluctuations of numerical density of states at the corners of the
region $[-3,3]$. In case of better numerical results (made for
much larger $N$´s) one could define precisely whether there are
more singularities of the derivative of $\rho(E)$ or not, choose
an appropriate solution of the singular integral equation and
organize better fit.

 {\bf Conclusions}: By use of technique developed in \cite{P-D,SS}
 we have calculated analytically exactly the
 Landauer resistance and the corresponding exponent in the random
 dimer model. We have analytically observed presence of delocalization
 transition for $m\leq 2$ confirming results obtained earlier. Corresponding
 critical indices have been found. For $m < 2$ we have obtained $\nu =2$.
 For limiting value $m =2$  we
 have obtained asymmetric behavior for the localization length
 when approaching to critical values $E_c= \pm 1$ from the left and right hand
 sides. It appeared, that $\nu = 1/2$ when approaching to boundaries
 of the segment $[-1,1]$ from outside and $\nu = 1$- from inside.
 It looks that this fact was not observed earlier.

 We compare the half of Landauer exponent with the numerical
 calculations of the Lyapunov exponent (Fig.1) and found good
 correspondence in the central region in accordance with
 \cite{ATAF}.

 By use of theory of singular integral equations \cite{M} we found
 analytic solution of the Thouless equation (\ref{Carleman}) for the density of
 states. Instead of Lyapunov exponent we put the half of Landauer exponent
into the expression of the density of states and compare it with
numerical calculations made by the diagonalization of the
Hamiltonian. As it was expected we got good correspondence ($1\%$)
in the central region. The fit of numerical results at the
boundaries can be improved by considering longer chain and
choosing solution of singular integral equations for multiply
connected regions.

{\bf Acknowledgment}: I am grateful to  A.Sedrakyan for the
formulation of the problem and support, P. Hasenfratz for many
valuable discussions and remarks and T.Hakobyan for the help in
numerical analyzes. I also would like to acknowledge the ITP of
Bern University for hospitality, where the major part of this work
was done.

This work was supported in part by  SNF SCOPE grant, Volkswagen
foundation and the INTAS grant 00561.

\pagestyle{plain} \makeatletter

\end{document}